\documentclass[review]{elsarticle}

\usepackage{lineno,hyperref}
\modulolinenumbers[1]

\usepackage{amsthm,amsmath}
\usepackage{caption}
\usepackage{subcaption}
\usepackage[utf8]{inputenc} 
\usepackage{graphics}
\usepackage{pgfplots}
\usepackage{tikz}

\usepackage{setspace}
\usepackage[margin=1in]{geometry}

\journal{Applied Soft Computing}

\begin{document}

\begin{frontmatter}

\title{Lightweight Speaker Verification for Online Identification of New Speakers with Short Segments}

\author[iimas]{Ivette V\'elez}
\author[iimas]{Caleb Rascon\corref{correspondingauthor}}
\cortext[correspondingauthor]{Corresponding author}
\ead{caleb.rascon@iimas.unam.mx}
\author[iimas]{Gibr\'an Fuentes-Pineda}

\address[iimas]{
  Instituto de Investigaciones en Matem\'aticas Aplicadas y en Sistemas (IIMAS), Universidad Nacional Aut\'onoma de M\'exico (UNAM), Circuito Escolar 3000, 04510, Mexico
  }

\begin{abstract}
\singlespacing

Verifying if two audio segments belong to the same speaker has been recently put forward as a flexible way to carry out speaker identification, since it does not require to be re-trained when new speakers appear on the auditory scene. Although many of the current techniques have achieved high performances, they require a considerably high amount of memory, and a specific minimum length for their input audio segments. These requirements limit the applicability of these techniques in scenarios such as service robots, internet of things and virtual assistants, where computational resources are limited and the users tend to speak in short segments. In this work we propose a BLSTM-based model that reaches a level of performance comparable to the current state of the art when using short input audio segments, while requiring a considerably less amount of memory. Further, as far as we know, a complete speaker identification system has not been reported using this verification paradigm. Thus, we present a complete online speaker identifier, based on a simple voting system, that shows that the proposed BLSTM-based model achieves a similar performance at identifying speakers online compared to the current state of the art.
\end{abstract}


\end{frontmatter}

\singlespacing

{
	\vspace{-1em}

	\small
	\noindent
	\textcopyright~2020. This manuscript version is made available under the \href{http://creativecommons.org/licenses/by-nc-nd/4.0/}{CC-BY-NC-ND 4.0 license}.

	\noindent
	Published article DOI: \url{https://doi.org/10.1016/j.asoc.2020.106704}
}


\section{Introduction} \label{sect:intro}

It is of great interest that computer systems interact with humans in a similar manner as a human would. Thus, there is a growing need to correctly identify the speaker by their voice alone \cite{Youssef2010}, and there has been recent important progress in terms of performance in the wild \cite{NAGRANI2020101027}. This progress has been largely based on the use of deep neural networks \citep{irum2019speaker}, which tend to occupy a considerable amount of computational resources, since the number of parameters used to obtain such a high performance is usually relatively high \cite{NAGRANI2020101027}. Additionally, several of these techniques tend to require a sizable segment of time with which to identify the user to obtain these high performances \cite{Hajavi2019}. These two requirements limit the application scenarios in which these high-performing speaker identification techniques can be used, such as service robots \cite{Grondin2012}, internet of things \cite{7224867}, and virtual assistants \cite{tiwari2018virtual}. In these scenarios, users speak in small spurts of time \cite{5453181}, requiring that the identification is carried out only using short segments of audio (between 0.5 and 1 s). Additionally, other processes are usually carried out in parallel (such as natural language processing, face recognition, action planning, etc.) and it is of interest that all functionalities are run on-site (in case of network outages). This limits the amount of memory and computational resources that can be used for speaker identification.

Although there has been an increasing amount of speaker identification techniques based on Convolutional Neural Networks (CNN), Bi-directional Long Short-Term Memory networks (BLSTM) have rarely been used for this purpose, while they have provided good results in other audio applications, such as voice conversion \cite{8695725}, sound source separation \cite{wang2018alternative} and speech recognition \cite{sym11050644}. An important aspect of BLSTM is their re-use of weights in their inner processes for modeling temporal data, which results in a smaller amount of parameters. Additionally, because of their recurrent nature, as well as their use of memory, they are well suited for finding temporal patterns. Both of these features in conjunction make them a viable candidate to explore for carrying out lightweight speaker verification.

In typical human-human interaction, new users are often introduced in the environment, such as when a new customer enters a restaurant or when a new house guest uses a device. Classical identification techniques rely either on a classification model, that has an output for every known speaker, or on a series of verification models, each trained to identify a known speaker  \cite{campbell1997}. Recently, there has been a shift away from this approach \cite{chung2019voxsrc} towards what we will refer to ``generic verification'', where a model is trained to compare two text-independent audio segments to establish if they are from the same user or not. This generality makes the solution space much more complex, but provides the benefit that, with an additional selection scheme, the two-input verification model can be used for speaker identification. It is important to state that, although generic verification can be used for this purpose, as far as we know there has not been reported a complete speaker identification system based on this paradigm.

In this work, we propose a BLSTM-based model to carry out generic speaker verification that requires a relatively small amount of parameters and short segments of audio. It is important to state that our proposal does not aim to outperform the current state of the art of speaker verification. It aims to offer a reasonable trade-off between performance and portability. Meaning, we believe that the differential of the computational and segment-length requirements between the proposed model and the current state-of-the-art heavily outweighs their performance differential. Additionally, we propose to use this model alongside a simple voting system, to provide a complete online speaker identifier that does not requires to be re-trained when new speakers are encountered.

For clarity, the originality of the proposed approach is the following:

\begin{itemize}
    \item Embedding-based speaker verification is carried out with only short segments of audio, while using a relatively small amount of computational requirements.
    \item A first approach to a complete generic-verification-based speaker identification is presented and evaluated.
\end{itemize}

To facilitate the adoption of our proposal, the source code and trained weights of the complete system can be freely downloaded from:

\centerline{\small \url{https://github.com/julik43/blstm_speaker_id}}

The remainder of this paper is organized as follows: a summary of works related to ours is presented in Section \ref{sect:related_work}; in Section \ref{sect:prop_system} the proposed BLSTM-based model is described; in Section \ref{sect:results}, the proposed BLSTM-model is evaluated and compared against the state of the art in terms of performance and memory usage; in Section \ref{sect:classif} a complete online speaker identifier is summarized, and is evaluated using the proposed BLSTM-based model and the state of the art; and, we conclude our work in Section \ref{sect:conclusions}.

\section{Related Work} \label{sect:related_work}

Speaker identification for a considerable amount of time has long been carried out by either classification or verification \cite{campbell1997}. However, recently there has been an important shift towards techniques that transform the input signal to a ``speaker domain'', where the speaker is represented by an embedding vector calculated from the input signal, and then compared against the embeddings of other known speakers. For example, in \cite{Daqrouq2011,DAQROUQ2015231} the embeddings were built from traditional features (formants and Shannon entropy wavelet packets), but a neural network was required to measure their similarity robustly. A more popular approach is based on i-vectors \cite{5545402}, but more recently the use of deep learning techniques have been more prominent for embedding calculation \cite{Snyder2017}. Convolutional Neural Networks (CNN) have been the more popular choice, since they have been well tested for feature extraction in computer vision. They have been used to generate new types of features which are then fed into different statistical methods \cite{Snyder2017,Snyder2016,Variani2014,Bhattacharya2017,Heigold2016}. Moreover, CNNs have also been used with raw audio \cite{Muckenhirn2017} to extract the relevant information to be used with an ad-hoc verifier generated for every speaker. They have also been extensively employed in a Siamese-fashion for biometric-based human identification for several years, e.g. in signature verification \cite{NIPS1993_769}, fingerprint recognition \cite{Baldi1993NeuralNF}, face verification (in conjunction with a similarity metric) \cite{Chopra05learninga}, and gait recognition \cite{zhangsnn2016}. These applications are compatible with Siamese networks since they can be used to verify if two input signals are from the same source (in these cases, from the same user).

To this effect, speaker identification is now being approached by ways of what we refer in this work as ``generic verification'': where a model is trained that establishes if two audio segments belong to the same speaker or not. In fact, the recent 2019 VoxCeleb Speaker Recognition Challenge (VoxSRC) \cite{chung2019voxsrc} established the goal of the contestants for this specific task. 

A representative example of this type of approach is the work of Nagrani et. al. \cite{nagrani17}, where the authors describe the VoxCeleb1 database and trained a Siamese CNN for generic verification of speakers. They use the cosine distance between two signals as a measure of similarity. For the identification process they report an accuracy of 80.5\% and 92.1\% for top-1 and top-5 respectively. Although the authors also report an identification, they did not use the generic verification paradigm to carry this out; they used a traditional classification approach. This work was extended to use a lightweight ``thin-ResNet'', with a NetVLAD-based time feature agreggator, that is able to estimate such embeddings from input segments with a variable length \cite{8683120}.

Interestingly, the vast majority of these works are based on the use of CNNs for embedding calculation. A rare exception is \cite{Mobiny2018}, where speaker verification is carried out by using a Siamese model of two Long Short-Term Memory (LSTM) networks, and a contrastive loss function used for verification. However, this approach involves the training of a verification model for each speaker, which is more akin to the classical verification approach. The authors reported an Equal Error Rate (EER) of 22.9\% and 22.1\% in their tests. As mentioned previously, since BLSTMs re-use weights in their inner processes for temporal modeling, they tend to employ a small amount of parameters.

Furthermore, even though the vast majority of the recent embedding-based techniques report impressive verification performances, they do not aim for ``lightweightedness''. Meaning, the amount of parameters they employ are usually quite high, limiting their applicability in scenarios such as service robotics, internet of things and virtual assistants. As for the length of audio segments, several seconds of information are required to obtain these high performances. A notable exception to this is the work of \cite{Hajavi2019}, where sub-1-second segments were tested with an EER below 7\% and memory usage was of 268 MB.

It is then of interest to have a speaker verification system that provides a trade-off between performance and computational and segment-length requirements.

It is important to note that, even though a possible use of embedding-based verification is for speaker identification, as far as we know, there has not been a report of a full speaker identification system based on this approach. To this effect, the work of \cite{Koch2015SiameseNN} approaches the task of classification of written characters by using embedding-based verification in conjunction with a simple voting-based selection scheme, and obtained good results. The same can be employed for speaker identification; and, as such, this approach is also explored in this work.

\section{Proposed BLSTM-Based Model} \label{sect:prop_system}

As described earlier, a recently popular approach for speaker identification is to train a system that establishes if two audio segments belong to the same speaker or not. This is carried out by calculating the embedding of the audio segments (to transform them into the ``speaker domain'') and then measuring their similarity. To calculate these embeddings, we propose a model based on a Bi-directional Long Short-Term Memory network (BLSTM), because of the relatively few amount of parameters that are employed to find temporal patterns. The aim is then to obtain a relatively good performance, using a relatively small amount of parameters and small input lengths.

\begin{figure}[ht]
    \centering
 	\includegraphics[angle=90,width=0.6\textwidth]{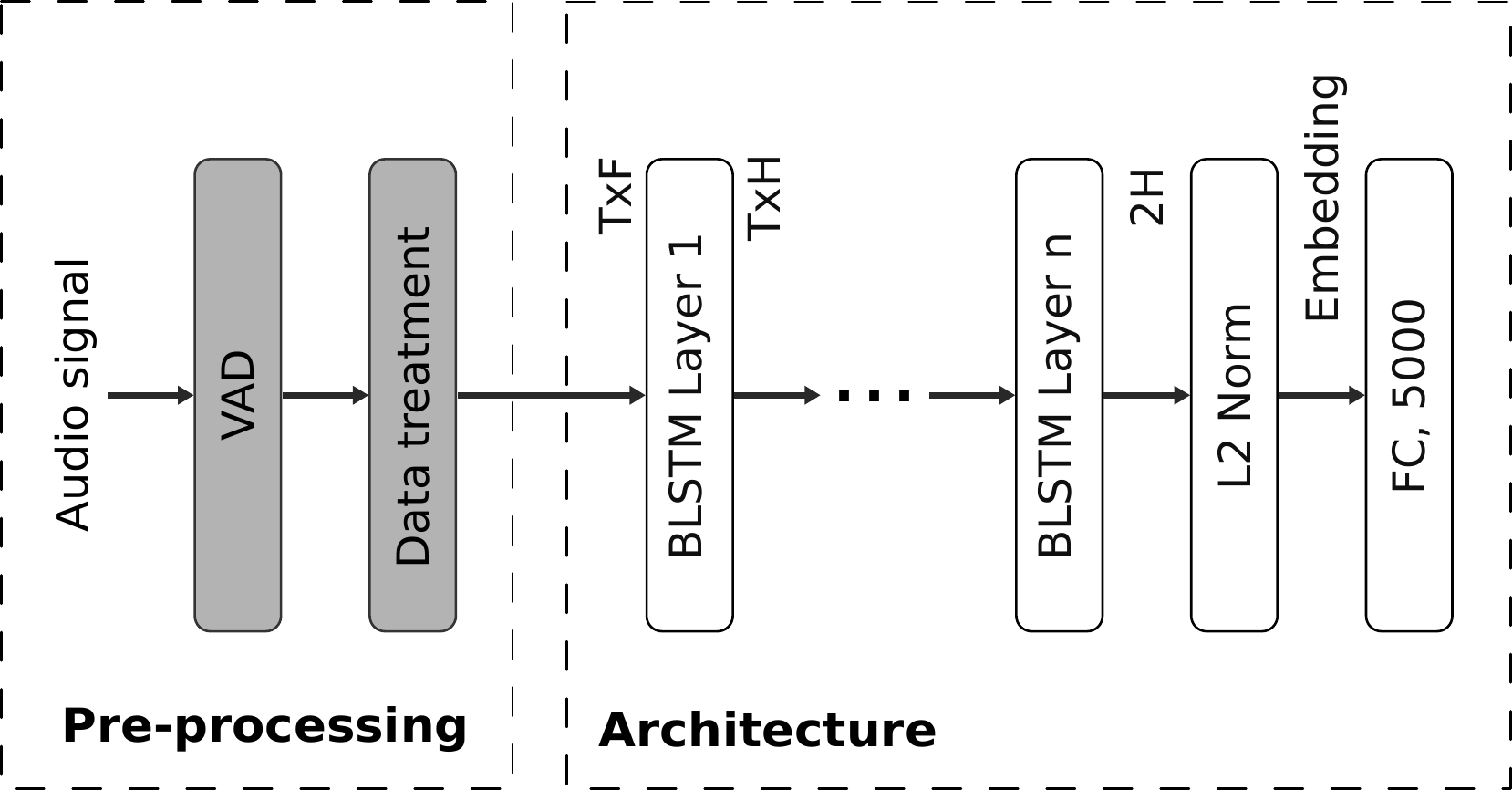}
	\caption{Proposed architecture for embedding calculation.}
	\label{fig:archtraining}       
\end{figure}

To train this model, we first establish a simple classification scenario, in which all but the last layer of the trained model is used for embedding calculation, as shown in Figure \ref{fig:archtraining}.

\clearpage

The last layer is a fully connected layer that carries out the classification from the embedding. This layer is then removed, and the rest of the network is then used for embedding calculation of incoming input segments. No fine-tuning is carried out afterwards.

The resulting network architecture is comprised of the three BLSTM layers with 256 units, and outputs an embedding of size 512, twice the number of units.

The similarity between the L2-normalized embeddings is calculated by their inner product.

For pre-processing, we employed the Voice Activity Detection technique based on \cite{vad}, which employs a 20 dB threshold to discriminate between silent and active windows. In terms of the input segment length, the model was trained with segments with lengths from 0.25 to 2.0 seconds.

In terms of employed features, we propose to explore different types of spectrograms, such that the input of the model is a matrix in which one dimension is time. In the other dimension, we extracted the following features to explore:

\begin{itemize}
	\item Spectral magnitude in a linear scale. Referred here as \textit{SpecMag}.
	\item Spectral magnitude in a decibel scale. The result is a type of frequency selector, since it tends to amplify frequency bins with high energy, while reducing ones with low energy.  Referred here as \textit{SpecdB}.
    \item Spectral density, estimated by the square of the linearly-scaled magnitude. The result provides an estimate of the energy distribution throughout the spectral range. Referred here as \textit{Spec}.
    \item Spectral magnitude in a linear scale after filtering the input audio with a simple pre-emphasis filter, which avoids distortion in high frequencies while reducing variability in the extracted spectra \cite{526613}. Referred here as \textit{EmphSpec}.
    \item The previous feature, but in a decibel scale. Referred here as \textit{EmphSpecdB}.
    \item The Mel-Frequency Cepstral Coefficient (MFCC) spectrum, built with 40 logarithmically spaced triangular filters. Referred here as \textit{MFCC}.
\end{itemize}

The time dimension of all the employed spectrograms were calculated by using either a Hamming window (for Spec, EmphSpec, EmphSpecdB and MFCC) or a Hann window (for SpecMag and SpecdB) of 32 ms, with an overlap of 16 ms. Given that the recordings used for training and validation (described in the following section) are sampled at 16 kHz, the length of the window is of 512 samples. An interesting note here is that, regardless of the input segment length, the embedding dimension will ultimately be of 512, which could be argued provides consistency in the embedding-domain search space during training.

\subsection{Training, Validation, and Testing Methodology} \label{subsec:train}

The VoxCeleb2 database \cite{voxceleb2} was used for training the classification network, from which the speaker embedding is calculated. This is the same training corpus that was used to train the model with which we compare our proposed system to (detailed in Section \ref{subsec:eervstime}). VoxCeleb2 has $1092009$ recordings of $5994$ speakers in its ``dev'' subset. To simplify the training process, we chose to use only $1000000$ randomly-selected recordings of only $5000$ randomly-chosen speakers of the same subset (as it provided an integer number of records used per speaker\footnote{This is preferable when using \textit{tfrecords} in TensorFlow in a multi-core system.}). It is important to mention, however, that the recordings per speaker were randomly selected at the start of each training epoch, thus, the whole utterances for each of the selected speakers of the VoxCeleb2 database were used to train our model. 

Each model was trained for 30 epochs using the Adam optimizer \cite{kingma2014adam}, a cross-entropy loss function, a learning rate of $0.0001$ and a batch size of $100$. After each training epoch, a validation stage was carried out with $8000$ randomly-selected entries to determine its classification performance during training. The amount of epochs was determined by previous tests that showed that the resulting embeddings provided the same or worse verification performance.

As mentioned before, a VAD system (which is based on \cite{vad}) is used to extract time segments of vocal activity from the corpus audio files.

For testing the trained model for generic verification, the trained network was evaluated with the VoxCeleb1 verification test list, released by the VGG group\footnote{Obtained from its GitHub repository in (\url{https://github.com/WeidiXie/VGG-Speaker-Recognition})}, composed of 37720 balanced data pairs of the VoxCeleb1 dataset. In this stage, for each audio in the test list, the embeddings of the pair of utterances were generated using the trained model, then the inner product was calculated between the pairs to determine if both were from the same speaker or not. This was carried out by using a threshold in the mid-point of $[-1,1]$ range of possible values of the inner product.

\section{Results} \label{sect:results}

Table \ref{table:verification_results} presents the memory usage (in MB) and the equal error rate (EER) of the verification of each of the models trained with every possible combination of segment lengths (0.25, 0.5, 1.5 and 2 s) and of explored input features (SpecdB, Spec, SpecMag, EmphSpec, EmphSpecdB and MFCC).

\begingroup
\renewcommand{\arraystretch}{0.75}
\begin{table}[ht]
\small
\centering
\begin{tabular}{rcccc}
    \hline
        Input Feat. & Length (s) & EER (\%) & Memory Usage (MB) \\ 
    \hline
        \textit{SpecdB} & \textit{2.00} & \textit{13.61} & \textit{16.80} \\ 
        Spec & 2.00 & 18.84 & 16.80 \\ 
        SpecMag & 2.00 & 14.22 & 16.80 \\ 
        EmphSpec & 2.00 & 14.11 & 16.80 \\ 
        EmphSpecdB & 2.00 & 13.53 & 16.80 \\ 
        MFCC & 2.00 & 13.63 & 15.03 \\ 
        \textit{SpecdB} & \textit{1.50} & \textit{14.81} & \textit{16.80} \\ 
        Spec & 1.50 & 20.79 & 16.80 \\ 
        SpecMag & 1.50 & 16.09 & 16.80 \\ 
        EmphSpec & 1.50 & 16.36 & 16.80 \\ 
        EmphSpecdB & 1.50 & 15.43 & 16.80 \\ 
        MFCC & 1.50 & 15.62 & 15.03 \\ 
        \textit{SpecdB} & \textit{1.00} & \textit{17.54} & \textit{16.80} \\ 
        Spec & 1.00 & 22.90 & 16.80 \\ 
        SpecMag & 1.00 & 19.09 & 16.80 \\ 
        EmphSpec & 1.00 & 19.03 & 16.80 \\ 
        EmphSpecdB & 1.00 & 18.61 & 16.80 \\ 
        MFCC & 1.00 & 18.07 & 15.03 \\ 
        \textit{SpecdB} & \textit{0.50} & \textit{24.84} & \textit{16.80} \\ 
        Spec & 0.50 & 29.41 & 16.80 \\ 
        SpecMag & 0.50 & 25.76 & 16.80 \\ 
        EmphSpec & 0.50 & 26.20 & 16.80 \\ 
        EmphSpecdB & 0.50 & 25.07 & 16.80 \\ 
        MFCC & 0.50 & 24.88 & 15.03 \\ 
        \textit{SpecdB} & \textit{0.25} & \textit{29.92} & \textit{16.80} \\ 
        Spec & 0.25 & 33.31 & 16.80 \\ 
        SpecMag & 0.25 & 30.96 & 16.80 \\ 
        EmphSpec & 0.25 & 31.64 & 16.80 \\ 
        EmphSpecdB & 0.25 & 30.00 & 16.80 \\ 
        MFCC & 0.25 & 29.79 & 15.03 \\ 
    \hline
\end{tabular}
\caption{Results of the evaluation of all trained models.}
\label{table:verification_results}
\end{table}
\endgroup

As it can be seen, for lengths 1.5 s and below, the SpecdB feature (a spectrogram with spectral magnitude in a decibel scale) and the MFCC feature consistently outperformed the other features in each possible segment length. Between these two features, the SpecdB feature outperformed the MFCC feature in more length scenarios. Because of these reasons, all of the following comparisons were carried out using the SpecdB feature as part of the proposed BLSTM-based model.

To fully evaluate our system we compare the BLSTM-based model to relevant state of the art models with varying degrees of input segment lengths, as well as memory usage.

\subsection{EER vs Input Segment Length}
\label{subsec:eervstime}

\begin{figure}[ht]
    \centering
 	\includegraphics[angle=90,width=0.68\textwidth]{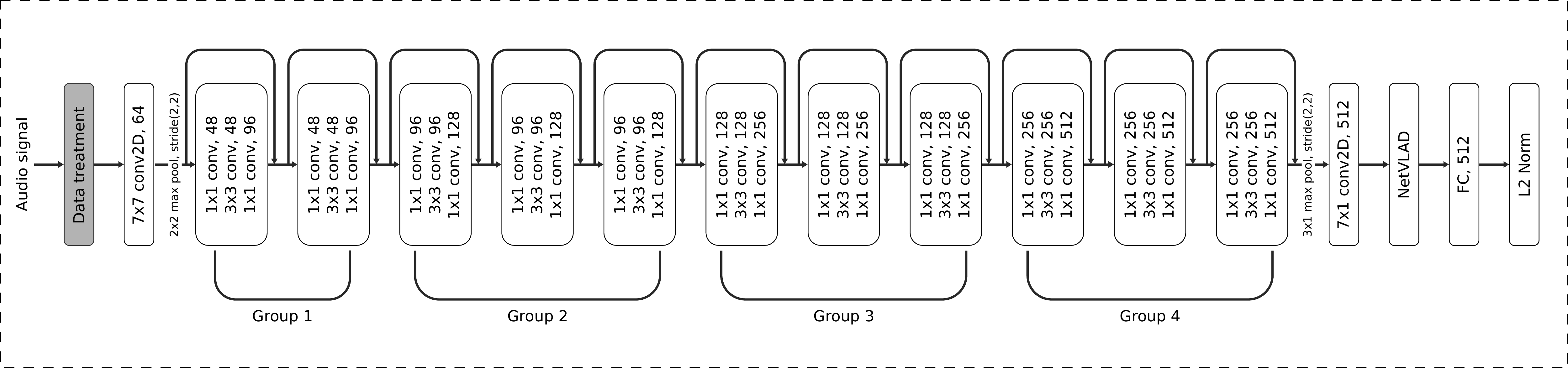}
	\caption{The network architecture presented in \cite{8683120}, referred here as \textit{thin-ResNet34}.}
	\label{fig:vgg}       
\end{figure}

In terms of what other embedding-based verification techniques with which to compare our system, i-vectors \cite{5545402}, x-vectors \cite{120006705553} and ResNet50-based \cite{he2015resnet}, as far as we know, have not been evaluated with small input segment lengths and are not publicly available. Thus, a direct comparison cannot be made. Thus, we chose the aforementioned work of \cite{8683120}, where the authors employed a lightweight `thin-ResNet'' with a NetVLAD-based aggregator to calculate embeddings, here referred to as \textit{thin-ResNet34}. It was chosen given that the model was publicly available, and is directly compatible with the comparison, since it supports input segments with variable time lengths without requiring to be re-trained. Additionally, it has shown good results with input lengths of 2 seconds and above. Thus, the high-performing thin-ResNet34 model is both publicly available (with trained weights) and is able to be evaluated with short segments, thus, it provides a good benchmark of the performance of the state of the art in the case scenario relevant to our application setting. Its architecture is shown in Figure \ref{fig:vgg} and no re-training was carried out for the results presented here.

\begin{figure}[ht]
    \centering
 	\includegraphics[width=0.55\textwidth]{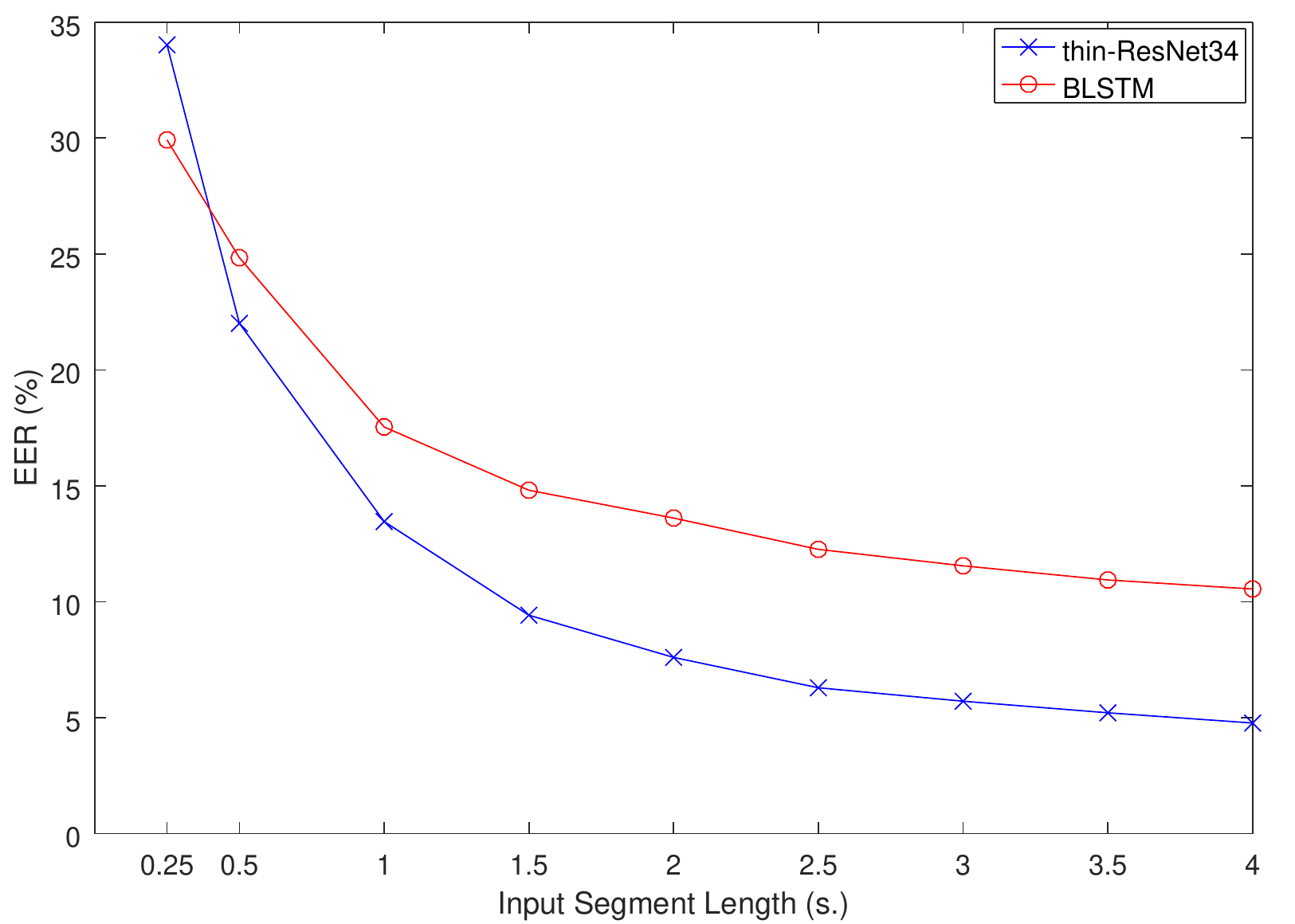}
	\caption{EER vs input time length using VoxCeleb1, with thin-ResNet34 and the BLSTM-based model.}
	\label{fig:comparison_results_voxceleb1}       
\end{figure}

The VoxCeleb1 corpus \cite{nagrani17} was used for evaluation, and the input pairs were selected in the same manner as described in \cite{8683120}. To ensure that our evaluation process does not deviate from previously reported results, we re-created the evaluation procedure used in \cite{8683120} and re-evaluated thin-ResNet34, and confirmed it reported the same results. Then, both thin-ResNet34 and the BLSTM-based model were evaluated using segments with lengths from 0.25 to 4 s, the results of which are shown in Figure \ref{fig:comparison_results_voxceleb1}.

It is clear that thin-ResNet34 provides lower EER than our system when using longer input segment lengths ($\geq 1$ s). However, both systems perform comparably with shorter segments.

\begin{figure}[ht]
    \centering
 	\includegraphics[width=0.55\textwidth]{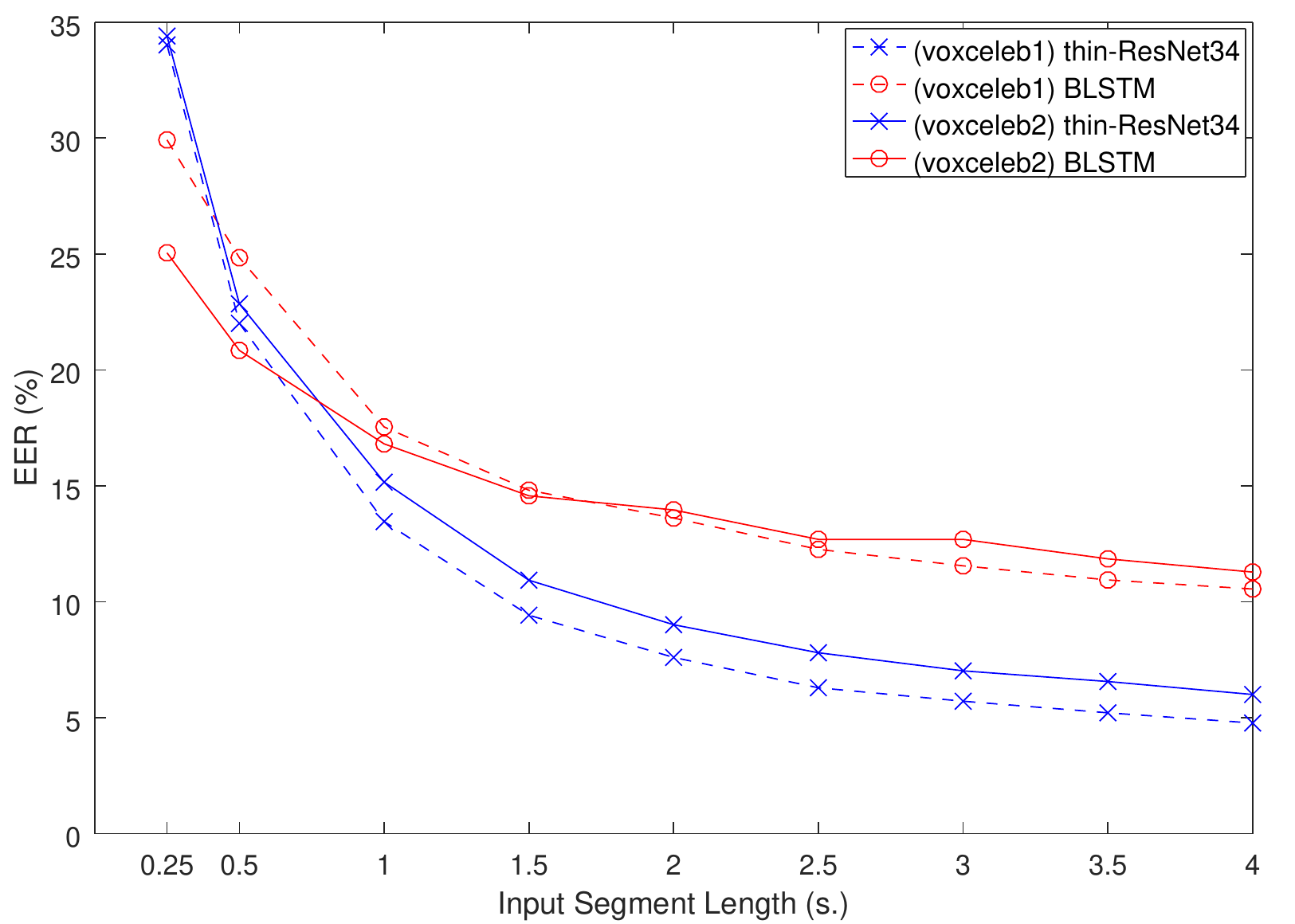}
	\caption{EER vs input time length using VoxCeleb2, with thin-ResNet34 and the BLSTM-based model, with the evaluation using VoxCeleb1 as comparison (dashed lines).}
	\label{fig:comparison_results_voxceleb2}
\end{figure}

It is also of interest to evaluate the consistency of the evaluated systems across different data sets. To this effect, the VoxCeleb2 corpus \cite{voxceleb2} was used to evaluate them both, using the same number of pairs of input segments (37720), randomly selected from the ``test'' subset. These results are shown in Figure \ref{fig:comparison_results_voxceleb2}, along with the results from Figure \ref{fig:comparison_results_voxceleb1} for comparison (as dashed lines).

As it can be seen, when tested with VoxCeleb2, our system also performs comparably to thin-ResNet34 with segments that are 1 s long and even outperforms it with segments $\leq 0.5$ s. Additionally, it is important to mention that the EER performance differential between both models in input segments with a length above 1 s does not surpass beyond 6 percentile points in any of the scenarios, which as it will be seen later, can be considered a good trade-off considering the memory usage reduction the proposed system provides.

\subsection{EER vs Memory Usage}

\begin{figure}[ht]
    \centering
 	\includegraphics[width=0.55\textwidth]{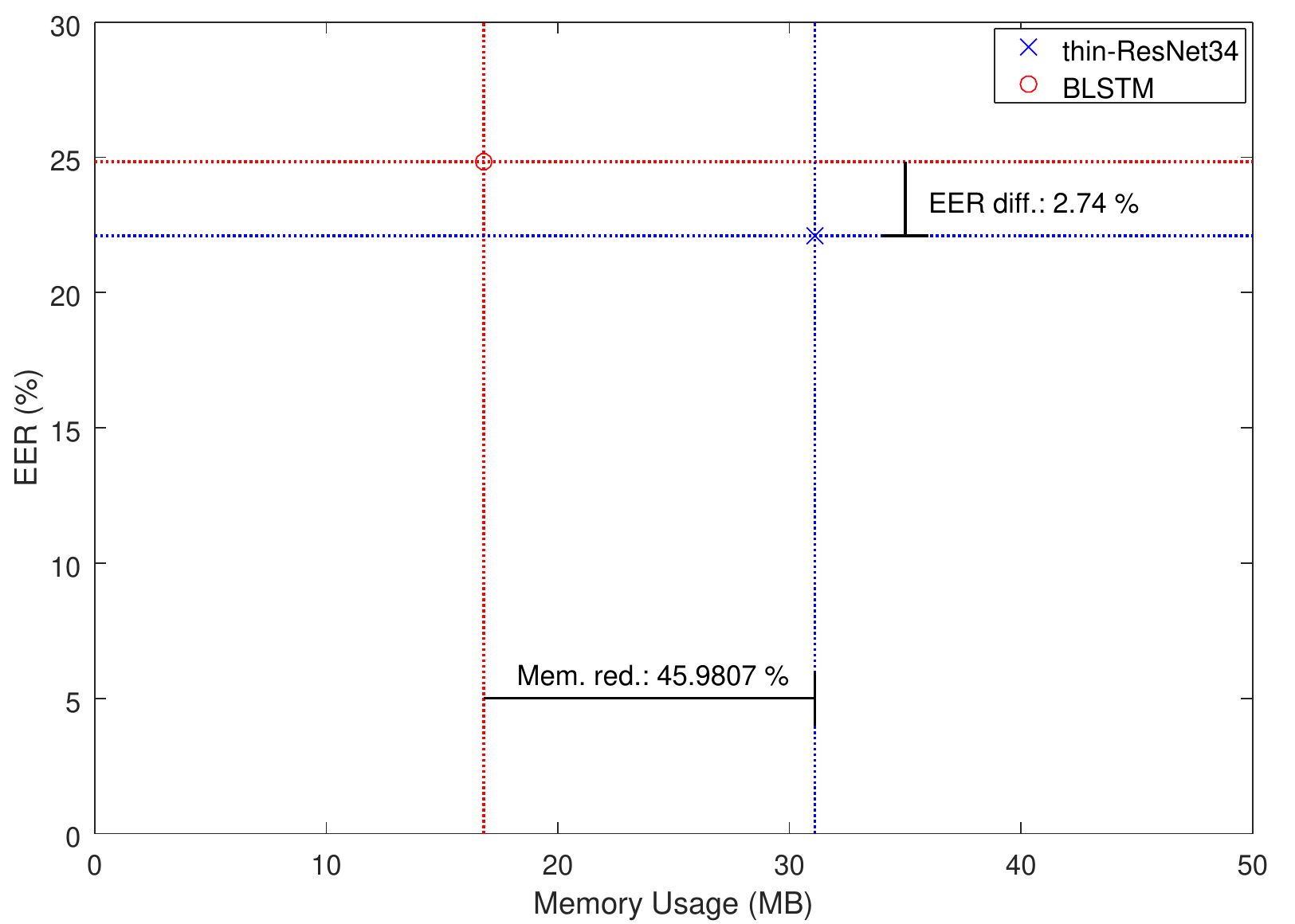}
	\caption{EER (input length of 0.5 s) vs memory usage with thin-ResNet34 and the BLSTM-based model using VoxCeleb1.}
	\label{fig:comparison_results_memory}
\end{figure}

It is also of interest to inspect the amount of memory used by the BLSTM-based model, and see if the loss in performance is a reasonable trade-off for lighter computational requirements and shorter input segment lengths. This comparison is shown in Figure \ref{fig:comparison_results_memory}, where for simplicity, the EER reported is the one obtained when using an input length of 0.5 s, when being evaluated with the VoxCeleb1 corpus. This input length was chosen because it has been found that in cases of interaction that have a high grade of back-and-forth between the user and the automatic conversational system, shorter utterances (between 0.5 and 1 s) are spoken more frequently by the human user \cite{5453181}.

As it can be seen, even though the EER performance differential is less than 3 percentile points using VoxCeleb1, the BLSTM-based model only uses nearly half the memory employed by thin-ResNet34.

\begin{figure}[ht]
    \centering
 	\includegraphics[width=0.55\textwidth]{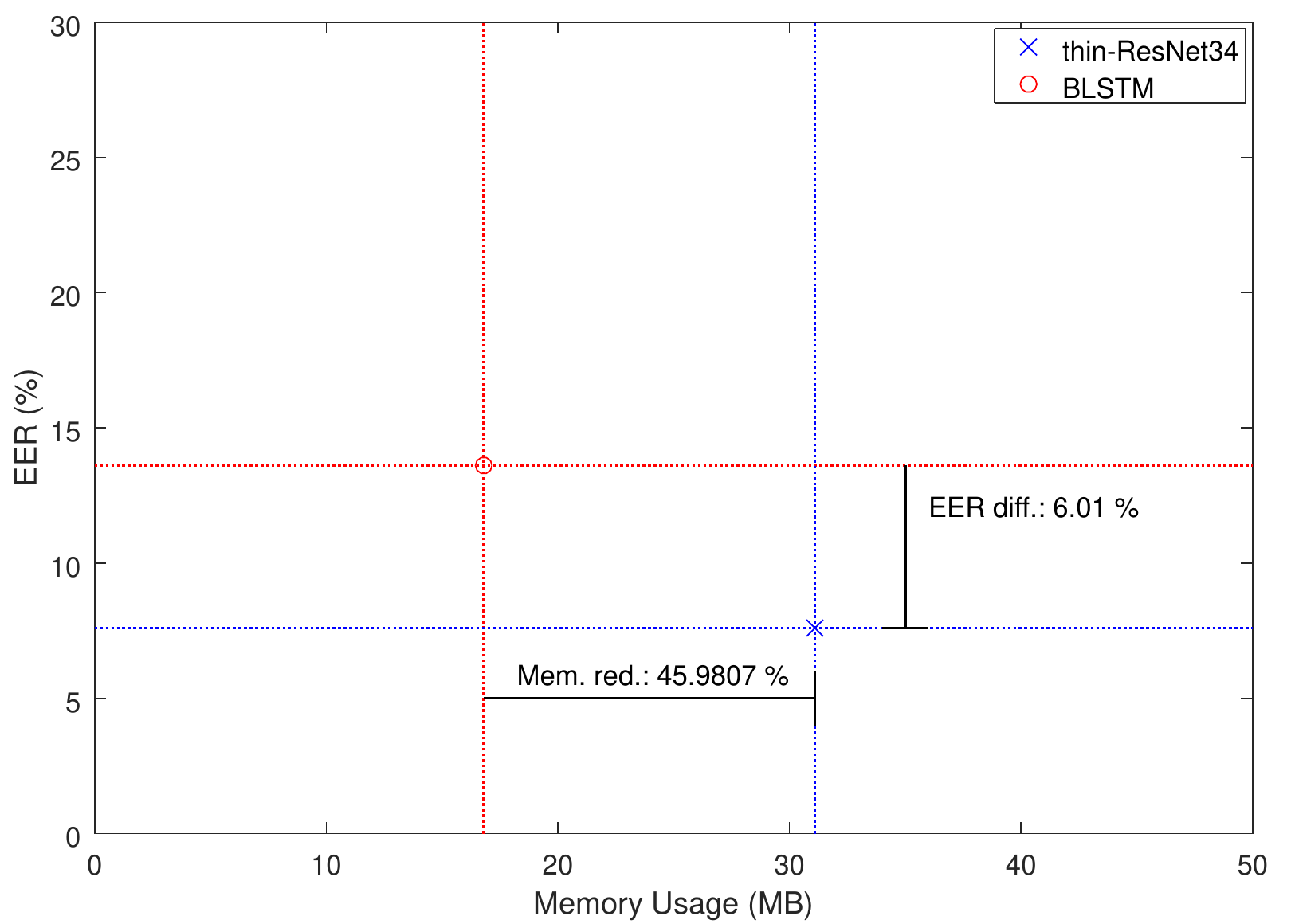}
	\caption{EER (input length of 2 s) vs memory usage with thin-ResNet34 and the BLSTM-based model using VoxCeleb1.}
	\label{fig:comparison_results_memory2}
\end{figure}

It is also of interests to carry out the same comparison with longer input segments. In Figure \ref{fig:comparison_results_memory2} the EER reported is the one obtained when using an input length of 2 s, when being evaluated with the VoxCeleb1 corpus. This input length was chosen since it had the highest inter-model EER difference.

As it can be seen, the EER performance differential only increased another 3 percentile points from the results shown in Figure \ref{fig:comparison_results_memory}. With a difference of 6 percentile points and a reduction of nearly half the memory employed, a case could be made that the trade-off is still worthwhile. It is important to re-iterate that the focus of the proposed system is with shorter input segments, and thus these evaluations with longer input segments are presented here as a type of characterization of the proposed system, and provided for the benefit of the reader to decide if such a trade-off is applicable in their case scenario.

For completeness sake, it is important to mention the work of UtterIdNet \cite{Hajavi2019}, which has achieved a very low EER of 6.46\% with input segments of 0.5 s length. However, as was reported in \cite{Hajavi2019}, UtterIdNet requires 268 MB of memory to run, an order of magnitude greater to the one required by the proposed BLSTM-model. To this effect, we believe that in applications such as service robots \cite{Grondin2012}, internet of things \cite{7224867}, and virtual assistants \cite{tiwari2018virtual}, the memory differential heavily outweighs the EER differential with shorter segments.

\section{Online Classification via a Voting System} \label{sect:classif}

To further compare the BLSTM-based model to current state of the art, we propose a complete online speaker identifier, based on a simple voting system. It is important to note that this proposal mainly serves as the basis of comparison between the two systems, and that more sophisticated voting systems may be applicable. However, we believe that it is important to report the results of an online speaker identifier (albeit a naive one) based on generic verification, so as to provide an initial baseline to the speaker identification community.

\begin{figure}[ht]
    \centering
 	\includegraphics[width=0.55\textwidth]{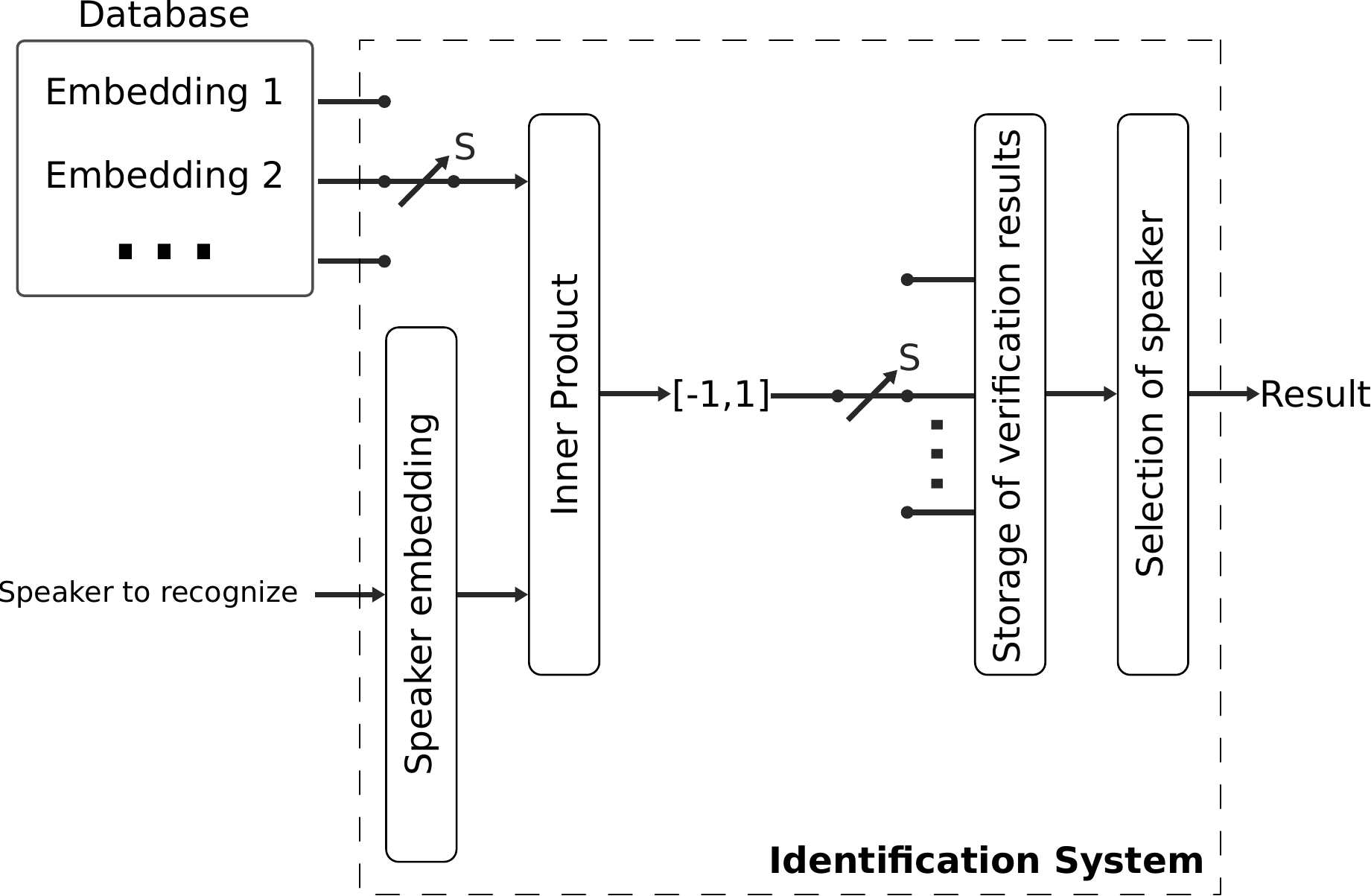}
	\caption{Diagram summarizing the complete online speaker identifier.}
	\label{fig:identification_diagram}
\end{figure}

The voting system is a selection scheme based on carrying out generic verification of the current audio input with each of the audio entries of an external database, each being an embedding vector belonging to a known speaker. A diagram of the whole identification process is shown in Figure \ref{fig:identification_diagram}.

\clearpage

To select the speaker to whom the audio input belongs to, the following steps are carried out:

\begin{enumerate}
	\item Calculate the embedding of the audio input.
	\item For each known speaker $c$, calculate the inner product to each of the audio entries that belong to $c$. Their average is the verification result $r_c$ of speaker $c$.
	\item Store in $\mathbf{R}$ all the calculated $r_c$.
	\item Apply Equation \ref{eq:1} to select the system's output $o$:
\end{enumerate}

\vspace{-10pt}

\begin{equation}\label{eq:1}
    o=\left\{\begin{matrix}
unknown, & if \quad \forall \quad r_{c}<0 \\
\underset{c}{\arg\max}(\mathbf{R}), & otherwise,
\end{matrix}\right.
\end{equation} 

Since the value of a verification result ranges between $[-1,1]$, the threshold of $0$ in Equation \ref{eq:1} is the mid-point of that range and, thus, provides a reasonable threshold to discern if the audio input belongs to a known user or not. If no member of $\mathbf{R}$ surpasses this threshold, the user that uttered the audio input is deemed unknown. If this is the case, a simple speech/keyboard interaction can be carried out to ask for the speaker's name, and subsequently add the embedding calculated from the audio input as an entry to the external database for their new class. If the speaker is deemed known, the embedding can be added to the external database as an additional entry for their class. 

\begin{figure}[ht]
	\centering
	
	\begin{subfigure}{0.4\textwidth}
		\centering
		\includegraphics[width=\textwidth]{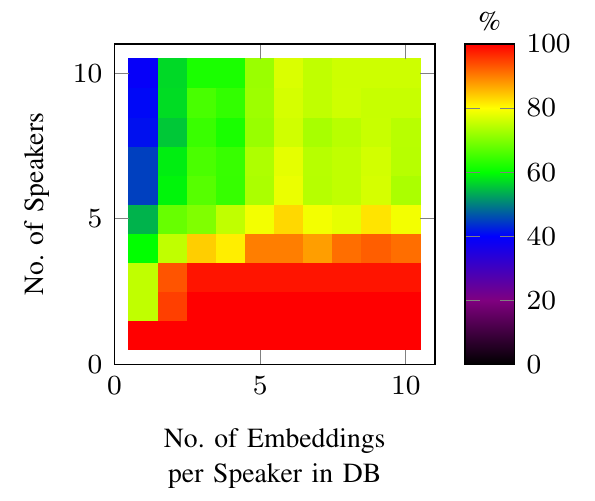}
		\caption{thin-ResNet34}
		\label{fig:heatmaps_vgg}
	\end{subfigure}
	\begin{subfigure}{0.4\textwidth}
		\centering
		\includegraphics[width=\textwidth]{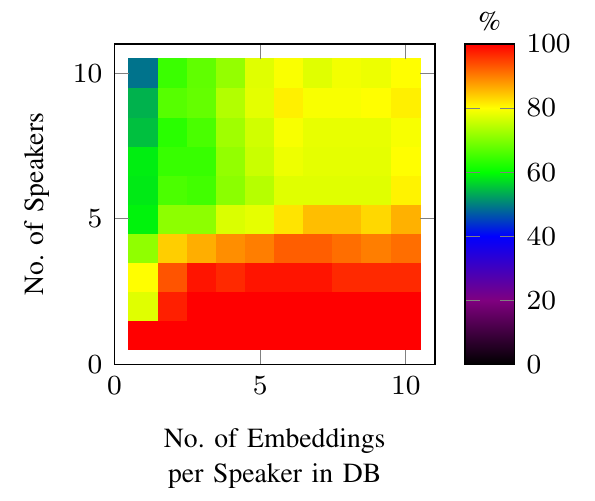}
		\caption{BLSTM-based model.}
		\label{fig:heatmap_proposed}       
	\end{subfigure}
	
	\caption{Accuracy heatmaps of the online speaker identifier using thin-ResNet34 and the BLSTM-based model for verification.}
	\label{fig:heatmaps}
\end{figure}

It is then of interest to evaluate this simple online speaker identifier when using the BLSTM-based model as well as thin-ResNet34 as its generic verifier. To this effect, an accuracy heatmap was created for each, where each cell in the heatmap represents a test configuration between a specific number of known speakers and a specific number of audio entries per speaker in the external database. Figure \ref{fig:heatmaps} shows both accuracy heatmaps.

As it can be seen, there is very little difference between both heatmaps, and the difference that does show relates to the BLSTM-based model slightly outperforming thin-ResNet34.

It is important to mention that this method is proposed as a first approach of a generic-verification-based speaker identification system. We propose as future work to explore other decision mechanisms such as: 1) stopping the search through $\mathbf{R}$ when a certain verification threshold is surpassed; 2) pre-ordering $\mathbf{R}$ based on Bayesian decision theory; or 3) employing meta learning methods commonly employed in one-shot and few-shot learning \cite{vinyals2016matching,snell2017prototypical,sung2018learning}.

\clearpage

\section{Conclusions} \label{sect:conclusions}

There has recently been a shift towards embedding-based generic speaker verification, with which online speaker identification can be carried out without requiring re-training when new speakers appear in the auditory scene. Impressive performances have been achieved by using CNN-based models, but they usually work well with large input segment lengths ($\geq$ 2 s.) and have considerably high computational requirements.

In this work, we proposed the use of a BLSTM-based model to calculate the embedding of the inputs, which provided performances comparable to the state of the art with shorter input segments, while requiring considerably less memory to achieve them.

Further, a complete online speaker identifier is presented, based on a simple voting scheme that uses generic verification to carry out speaker identification without requiring to be re-trained with new speakers. The identifier was evaluated with the BLSTM-based model and the state of the art, and different testing configurations were carried with both, in which different amounts of known speakers were tested with different amounts of entries per known speaker. The accuracy was very similar throughout all of the different testing configurations when using a short input segment (0.5 s), while only using half of the memory that the state of the art employs.

For future work, other decision mechanisms will be explored to increase the accuracy of the online speaker identification system and stricter rules will be tested for database management to increase robustness while maintaining low response times.

\section*{Acknowledgements}
This work was supported by CONACYT grant [251319], and PAPIIT grants [IA100129] and [IA104016]. The authors would also like to thank Alejandro Maldonado for his support in code reviewing.

\section*{Competing interests}
The authors declare that they have no competing interests.

\section*{Availability of data and materials}
The code of the complete system, as well as the trained weights, can be found in \url{https://github.com/julik43/blstm_speaker_id}.

\bibliography{myrefs}

\begin{thebibliography}{10}
\expandafter\ifx\csname url\endcsname\relax
  \def\url#1{\texttt{#1}}\fi
\expandafter\ifx\csname urlprefix\endcsname\relax\def\urlprefix{URL }\fi
\expandafter\ifx\csname href\endcsname\relax
  \def\href#1#2{#2} \def\path#1{#1}\fi

\bibitem{Youssef2010}
K.~Youssef, S.~Argentieri, J.~Zarader, Binaural speaker recognition for
  humanoid robots, in: 2010 11th International Conference on Control Automation
  Robotics Vision, 2010, pp. 2295--2300.
\newblock \href {http://dx.doi.org/10.1109/ICARCV.2010.5707878}
  {\path{doi:10.1109/ICARCV.2010.5707878}}.

\bibitem{NAGRANI2020101027}
A.~Nagrani, J.~S. Chung, W.~Xie, A.~Zisserman, Voxceleb: Large-scale speaker
  verification in the wild, Computer Speech and Language 60.
\newblock \href {http://dx.doi.org/10.1016/j.csl.2019.101027}
  {\path{doi:10.1016/j.csl.2019.101027}}.

\bibitem{irum2019speaker}
A.~Irum, A.~Salman, Speaker verification using deep neural networks: A review,
  International Journal of Machine Learning and Computing 9~(1).

\bibitem{Hajavi2019}
A.~Hajavi, A.~Etemad, {A Deep Neural Network for Short-Segment Speaker
  Recognition}, in: Proc. Interspeech 2019, 2019, pp. 2878--2882.
\newblock \href {http://dx.doi.org/10.21437/Interspeech.2019-2240}
  {\path{doi:10.21437/Interspeech.2019-2240}}.

\bibitem{Grondin2012}
F.~Grondin, F.~Michaud, Wiss, a speaker identification system for mobile
  robots, in: 2012 IEEE International Conference on Robotics and Automation,
  2012, pp. 1817--1822.
\newblock \href {http://dx.doi.org/10.1109/ICRA.2012.6224729}
  {\path{doi:10.1109/ICRA.2012.6224729}}.

\bibitem{7224867}
D.~{Shin}, M.~{Jun}, Home iot device certification through speaker recognition,
  in: 2015 17th International Conference on Advanced Communication Technology
  (ICACT), 2015, pp. 600--603.
\newblock \href {http://dx.doi.org/10.1109/ICACT.2015.7224867}
  {\path{doi:10.1109/ICACT.2015.7224867}}.

\bibitem{tiwari2018virtual}
V.~Tiwari, M.~F. Hashmi, A.~Keskar, N.~Shivaprakash, Virtual home assistant for
  voice based controlling and scheduling with short speech speaker
  identification, Multimedia Tools and Applications (2018) 1--26.

\bibitem{5453181}
C.~{Yu}, M.~{Scheutz}, P.~{Schermerhorn}, Investigating multimodal real-time
  patterns of joint attention in an hri word learning task, in: 2010 5th
  ACM/IEEE International Conference on Human-Robot Interaction (HRI), 2010, pp.
  309--316.
\newblock \href {http://dx.doi.org/10.1109/HRI.2010.5453181}
  {\path{doi:10.1109/HRI.2010.5453181}}.

\bibitem{8695725}
X.~{Miao}, X.~{Zhang}, M.~{Sun}, C.~{Zheng}, T.~{Cao}, A blstm and
  wavenet-based voice conversion method with waveform collapse suppression by
  post-processing, IEEE Access 7 (2019) 54321--54329.
\newblock \href {http://dx.doi.org/10.1109/ACCESS.2019.2912926}
  {\path{doi:10.1109/ACCESS.2019.2912926}}.

\bibitem{wang2018alternative}
Z.-Q. Wang, J.~Le~Roux, J.~R. Hershey, Alternative objective functions for deep
  clustering, in: 2018 IEEE International Conference on Acoustics, Speech and
  Signal Processing (ICASSP), IEEE, 2018, pp. 686--690.

\bibitem{sym11050644}
D.~Wang, X.~Wang, S.~Lv, End-to-end mandarin speech recognition combining cnn
  and blstm, Symmetry 11~(5).
\newblock \href {http://dx.doi.org/10.3390/sym11050644}
  {\path{doi:10.3390/sym11050644}}.

\bibitem{campbell1997}
J.~P. {Campbell}, Speaker recognition: a tutorial, Proceedings of the IEEE
  85~(9) (1997) 1437--1462.
\newblock \href {http://dx.doi.org/10.1109/5.628714}
  {\path{doi:10.1109/5.628714}}.

\bibitem{chung2019voxsrc}
J.~S. Chung, A.~Nagrani, E.~Coto, W.~Xie, M.~McLaren, D.~A. Reynolds,
  A.~Zisserman, Voxsrc 2019: The first voxceleb speaker recognition challenge
  (2019).
\newblock \href {http://arxiv.org/abs/1912.02522} {\path{arXiv:1912.02522}}.

\bibitem{Daqrouq2011}
K.~Daqrouq, Wavelet entropy and neural network for text-independent speaker
  identification, Engineering Applications of Artificial Intelligence 24~(5)
  (2011) 796 -- 802.
\newblock \href {http://dx.doi.org/10.1016/j.engappai.2011.01.001}
  {\path{doi:10.1016/j.engappai.2011.01.001}}.

\bibitem{DAQROUQ2015231}
K.~Daqrouq, T.~A. Tutunji, Speaker identification using vowels features through
  a combined method of formants, wavelets, and neural network classifiers,
  Applied Soft Computing 27 (2015) 231 -- 239.
\newblock \href {http://dx.doi.org/10.1016/j.asoc.2014.11.016}
  {\path{doi:10.1016/j.asoc.2014.11.016}}.

\bibitem{5545402}
N.~{Dehak}, P.~J. {Kenny}, R.~{Dehak}, P.~{Dumouchel}, P.~{Ouellet}, Front-end
  factor analysis for speaker verification, IEEE Transactions on Audio, Speech,
  and Language Processing 19~(4) (2011) 788--798.
\newblock \href {http://dx.doi.org/10.1109/TASL.2010.2064307}
  {\path{doi:10.1109/TASL.2010.2064307}}.

\bibitem{Snyder2017}
D.~Snyder, D.~Garcia-Romero, D.~Povey, S.~Khudanpur, Deep neural network
  embeddings for text-independent speaker verification, in: Proc. Interspeech,
  2017, pp. 999--1003.
\newblock \href {http://dx.doi.org/10.21437/Interspeech.2017-620}
  {\path{doi:10.21437/Interspeech.2017-620}}.

\bibitem{Snyder2016}
D.~Snyder, P.~Ghahremani, D.~Povey, D.~Garcia-Romero, Y.~Carmiel, S.~Khudanpur,
  Deep neural network-based speaker embeddings for end-to-end speaker
  verification, in: IEEE Spoken Language Technology Workshop (SLT), 2016, pp.
  165--170.
\newblock \href {http://dx.doi.org/10.1109/SLT.2016.7846260}
  {\path{doi:10.1109/SLT.2016.7846260}}.

\bibitem{Variani2014}
E.~Variani, X.~Lei, E.~McDermott, I.~L. Moreno, J.~Gonzalez-Dominguez, Deep
  neural networks for small footprint text-dependent speaker verification, in:
  2014 IEEE International Conference on Acoustics, Speech and Signal Processing
  (ICASSP), 2014, pp. 4052--4056.
\newblock \href {http://dx.doi.org/10.1109/ICASSP.2014.6854363}
  {\path{doi:10.1109/ICASSP.2014.6854363}}.

\bibitem{Bhattacharya2017}
G.~Bhattacharya, J.~Alam, P.~Kenny, Deep speaker embeddings for short-duration
  speaker verification, in: Proc. Interspeech, 2017, pp. 1517--1521.
\newblock \href {http://dx.doi.org/10.21437/Interspeech.2017-1575}
  {\path{doi:10.21437/Interspeech.2017-1575}}.

\bibitem{Heigold2016}
G.~Heigold, I.~Moreno, S.~Bengio, N.~Shazeer, End-to-end text-dependent speaker
  verification, in: 2016 IEEE International Conference on Acoustics, Speech and
  Signal Processing (ICASSP), 2016, pp. 5115--5119.
\newblock \href {http://dx.doi.org/10.1109/ICASSP.2016.7472652}
  {\path{doi:10.1109/ICASSP.2016.7472652}}.

\bibitem{Muckenhirn2017}
H.~{Muckenhirn}, M.~{Magimai.-Doss}, S.~{Marcell}, Towards directly modeling
  raw speech signal for speaker verification using cnns, in: 2018 IEEE
  International Conference on Acoustics, Speech and Signal Processing (ICASSP),
  2018, pp. 4884--4888.
\newblock \href {http://dx.doi.org/10.1109/ICASSP.2018.8462165}
  {\path{doi:10.1109/ICASSP.2018.8462165}}.

\bibitem{NIPS1993_769}
J.~Bromley, J.~Bentz, L.~Bottou, I.~Guyon, Y.~LeCun, C.~Moore, E.~Sackinger,
  R.~Shah, Signature verification using a siamese time delay neural network,
  International Journal of Pattern Recognition and Artificial Intelligence
  7~(4).

\bibitem{Baldi1993NeuralNF}
P.~Baldi, Y.~Chauvin, Neural networks for fingerprint recognition, Neural
  Computation 5 (1993) 402--418.
\newblock \href {http://dx.doi.org/10.1162/neco.1993.5.3.402}
  {\path{doi:10.1162/neco.1993.5.3.402}}.

\bibitem{Chopra05learninga}
S.~{Chopra}, R.~{Hadsell}, Y.~{LeCun}, Learning a similarity metric
  discriminatively, with application to face verification, in: 2005 IEEE
  Computer Society Conference on Computer Vision and Pattern Recognition
  (CVPR'05), Vol.~1, 2005, pp. 539--546.
\newblock \href {http://dx.doi.org/10.1109/CVPR.2005.202}
  {\path{doi:10.1109/CVPR.2005.202}}.

\bibitem{zhangsnn2016}
C.~{Zhang}, W.~{Liu}, H.~{Ma}, H.~{Fu}, Siamese neural network based gait
  recognition for human identification, in: 2016 IEEE International Conference
  on Acoustics, Speech and Signal Processing (ICASSP), 2016, pp. 2832--2836.
\newblock \href {http://dx.doi.org/10.1109/ICASSP.2016.7472194}
  {\path{doi:10.1109/ICASSP.2016.7472194}}.

\bibitem{nagrani17}
A.~Nagrani, J.~S. Chung, A.~Zisserman, Voxceleb: A large-scale speaker
  identification dataset, in: Proc. Interspeech, 2017, pp. 2616--2620.
\newblock \href {http://dx.doi.org/10.21437/Interspeech.2017-950}
  {\path{doi:10.21437/Interspeech.2017-950}}.

\bibitem{8683120}
W.~{Xie}, A.~{Nagrani}, J.~S. {Chung}, A.~{Zisserman}, Utterance-level
  aggregation for speaker recognition in the wild, in: ICASSP 2019 - 2019 IEEE
  International Conference on Acoustics, Speech and Signal Processing (ICASSP),
  2019, pp. 5791--5795.
\newblock \href {http://dx.doi.org/10.1109/ICASSP.2019.8683120}
  {\path{doi:10.1109/ICASSP.2019.8683120}}.

\bibitem{Mobiny2018}
A.~Mobiny, Text-independent speaker verification using long short-term memory
  networks, ArXiv abs/1805.00604.

\bibitem{Koch2015SiameseNN}
G.~Koch, R.~Zemel, R.~Salakhutdinov, Siamese neural networks for one-shot image
  recognition, in: IMCL, Deep Learning Workshop, 2015, pp. 1--8.

\bibitem{vad}
catedrago, Audio split, \url{https://github.com/catedrago/split\_audio\_VAD}
  (2018).

\bibitem{526613}
R.~{Vergin}, D.~{O'Shaughnessy}, Pre-emphasis and speech recognition, in:
  Proceedings 1995 Canadian Conference on Electrical and Computer Engineering,
  Vol.~2, 1995, pp. 1062--1065 vol.2.
\newblock \href {http://dx.doi.org/10.1109/CCECE.1995.526613}
  {\path{doi:10.1109/CCECE.1995.526613}}.

\bibitem{voxceleb2}
J.~S. Chung, A.~Nagrani, A.~Zisserman, Voxceleb2: Deep speaker recognition, in:
  Proc. Interspeech, 2018, pp. 1086--1090.
\newblock \href {http://dx.doi.org/10.21437/Interspeech.2018-1929}
  {\path{doi:10.21437/Interspeech.2018-1929}}.

\bibitem{kingma2014adam}
D.~P. Kingma, J.~Ba, Adam: A method for stochastic optimization, arXiv preprint
  arXiv:1412.6980.

\bibitem{120006705553}
K.~Okabe, T.~Koshinaka, K.~Shinoda, Attentive statistics pooling for deep
  speaker embedding, in: Proc. Interspeech, 2018, pp. 2252--2256.
\newblock \href {http://dx.doi.org/10.21437/Interspeech.2018-993}
  {\path{doi:10.21437/Interspeech.2018-993}}.

\bibitem{he2015resnet}
K.~{He}, X.~{Zhang}, S.~{Ren}, J.~{Sun}, Deep residual learning for image
  recognition, in: 2016 IEEE Conference on Computer Vision and Pattern
  Recognition (CVPR), 2016, pp. 770--778.
\newblock \href {http://dx.doi.org/10.1109/CVPR.2016.90}
  {\path{doi:10.1109/CVPR.2016.90}}.

\bibitem{vinyals2016matching}
O.~Vinyals, C.~Blundell, T.~Lillicrap, D.~Wierstra, et~al., Matching networks
  for one shot learning, in: Advances in neural information processing systems,
  2016, pp. 3630--3638.

\bibitem{snell2017prototypical}
J.~Snell, K.~Swersky, R.~Zemel, Prototypical networks for few-shot learning,
  in: Advances in neural information processing systems, 2017, pp. 4077--4087.

\bibitem{sung2018learning}
F.~Sung, Y.~Yang, L.~Zhang, T.~Xiang, P.~H. Torr, T.~M. Hospedales, Learning to
  compare: Relation network for few-shot learning, in: Proceedings of the IEEE
  Conference on Computer Vision and Pattern Recognition, 2018, pp. 1199--1208.

\end{thebibliography}

\end{document}